%% file: extended-abstract.tex
\documentclass{sigchi-ext}
\usepackage[T1]{fontenc}
\usepackage{textcomp}
\usepackage[scaled=.92]{helvet} 
\usepackage{graphicx} 
\usepackage{balance}  
\usepackage{booktabs} 
\usepackage{ccicons}  
\usepackage{ragged2e} 

\def\plaintitle{SIGCHI Extended Abstracts Sample File: Note Initial
  Caps} 
\def\emptyauthor{}
\def\plainkeywords{Authors' choice; of terms; separated; by
  semicolons; include commas, within terms only; required.}

\title{A Case-Study on Variations Observed in Accelerometers Across Devices.}

\numberofauthors{5}

\author{%
  \alignauthor{%
   \textbf{Carlos Alvarado}\\
    \affaddr{University of South Florida}\\
    \affaddr{Tampa, FL 33620, USA}\\
    \email{calvarado5@mail.usf.edu} }  \alignauthor{%
    \textbf{Ghulam Jilani Quadri}\\
    \affaddr{University of South Florida} \\
    \affaddr{Tampa, FL 33620, USA} \\
    \email{ghulamjilani@usf.edu} } \vfil
    \alignauthor{%
    \textbf{Jennifer Adorno Nieves}\\
    \affaddr{University of South Florida}\\
    \affaddr{Tampa, FL 33620, USA}\\
    \email{jorgea1@usf.edu} }
    \alignauthor{%
    \textbf{Paul Rosen}\\    
    \affaddr{University of South Florida}\\
    \affaddr{Tampa, FL 33620, USA}\\
    \email{prosen@usf.edu} }
    }

\definecolor{linkColor}{RGB}{6,125,233}
\hypersetup{%
  pdftitle={\plaintitle},
  pdfauthor={\emptyauthor},
  pdfkeywords={\plainkeywords},
  bookmarksnumbered,
  pdfstartview={FitH},
  colorlinks,
  citecolor=black,
  filecolor=black,
  linkcolor=black,
  urlcolor=linkColor,
  breaklinks=true,
}

\begin{document}

\CopyrightYear{2020}
\setcopyright{rightsretained}

\copyrightinfo{Permission to make digital or hard copies of part or all of this work for personal or
classroom use is granted without fee provided that copies are not made or distributed
for profit or commercial advantage and that copies bear this notice and the full citation
on the first page. Copyrights for third-party components of this work must be honored.
For all other uses, contact the owner/author(s). \\ \vspace{4pt} Copyright held by the owner/author(s).}

\maketitle

\RaggedRight{} 

\begin{abstract}

Every year we grow more dependent on wearable devices to gather personalized data, such as our movements, heart rate, respiration, etc. To capture this data, devices contain sensors, such as accelerometers and gyroscopes, that are able to measure changes in their surroundings and pass along the information for better informed decisions. Although these sensors should behave similarly in different devices, that is not always the case. In this case study, we analyze accelerometers from three different devices recording the same actions with an aim to determine whether the discrepancies are due to variability within or between devices. We found the most significant variation between devices with different specifications, such as sensitivity and sampling frequency. Nevertheless, variance in the data should be assumed, even if data is gathered from the same person, activity, and type of sensor.
\end{abstract}




\section{Introduction}
\input{Intro.tex}

\section{Methods}
\input{Method.tex}

\section{Findings}

\input{Findings.tex}

\section{Discussion}
\input{Discussion.tex}

\section{Conclusion and What's Next}
\input{Conclusion.tex}

\section{Acknowledgements}
This project is supported in part by the National Science Foundation (IIS-1845204 and CNS-1458928, an REU Site on Ubiquitous Sensing).

\balance{} 

\bibliographystyle{SIGCHI-Reference-Format}
\bibliography{refs}

\end{document}

%% file: Intro.tex
The use of wearable devices to monitor body signals has led to improvements in mobile healthcare applications over recent decades~\cite{bariya2018wearable, 8796629, zheng2014unobtrusive}. Nowadays, these sensors are integrated into consumer products, such as smartwatches that monitor heart rate and movement~\cite{ishikawa2018find, king2018survey, reeder2016health}. 

Although we have come to depend on these sensors, not all are created equal~\cite{caulfield2019not}. Factors such as alignment, calibration, orientation, and precision can cause algorithms to reach incorrect determinations. Furthermore, sensors can become defective as they age and can lead to erroneous information being forwarded to an algorithm. While sensors of the same type are expected to record the same information, this is not always the case~\cite{greis2018uncertainty}. In some applications, such as medical settings, incorrect determinations can lead to incorrect interventions that have negative impacts on a patient's health.

It is our hypothesis that to improve an algorithm's performance and transferability, we must measure and understand the variability produced within and between different devices. To achieve this, we identify and investigate irregularities in sensor data recorded within a controlled setting.

In order to better detect these discrepancies, anomaly detection in the time-series datasets can be performed through various signal processing techniques, as well as through visualizing sensor data to facilitate finding patterns, connections, and missing records. 
Visualizing sensor data is nontrivial because of its size, complexity, and the high precision needed.
Significant effort has been applied towards the design of better visualizations for time-series and sensor data~\cite{albers2014task,chung2004real, mori2007typical, quadri2017visual, richter2009visualizing, walker2015timenotes}.

\begin{margintable}[-9pc]
  \begin{minipage}{\marginparwidth}
    \centering
    \begin{tabular}{l r}
      & {\small\textbf{Device}} \\
      & {\small\textbf{Specification}} \\
      \toprule
      Shimmer3 &g = 16g, \\ & f = 102.4Hz \\ 
      \midrule
      Z2 Force &g = 8g, \\ & f = 400Hz \\ 
      \midrule
      Nexus 5  &g = 2g,\\ & f = 200Hz \\ 
      \bottomrule
    \end{tabular}
    \caption{Device specification used in the case-study.
    g: max gravity force; f: sampling rate
    }
    \label{tab:table2}
  \end{minipage}
\end{margintable}

For this case study we evaluate accelerometers, due in part to their wide-spread use. In addition to medical applications, accelerometers are used for a variety of interactions with computing devices, and any irregularities or variance in the data measurements may lead to incorrect interactions and negative experiences.

%% file: Method.tex
An accelerometer is a sensor used to measure the acceleration force applied to a device~\cite{lara2012survey}. They are found on everyday devices, such as smartphones and fitness bands, where they control screen orientations and measure activity throughout the day. Even when devices are stationary, accelerometers are able to detect forces in our environment, such as gravity. As manufacturing technology has evolved over recent years, accelerometers have become more precise, but the additional precision can cause data to become noisy, as even the faintest of forces is detected.
Since these sensors can add noise and produce faulty data, it is important to determine when they are relaying incorrect signal information.

In this work, we measure the differences captured by various accelerometers when the same amount of force is applied. To achieve this, we had participants perform various activities, while multiple accelerometers were worn simultaneously with the same location and alignment. These sessions were recorded and synchronized to measure variability between them.

\subsection{\textbf{Data Collection}}

The signals generated by two or more sensors differ because of inherent variability, which is forwarded to algorithms, which may result in incorrect conclusions.

\begin{marginfigure}[4pc]
    \centering
    \includegraphics[width=0.98\marginparwidth]{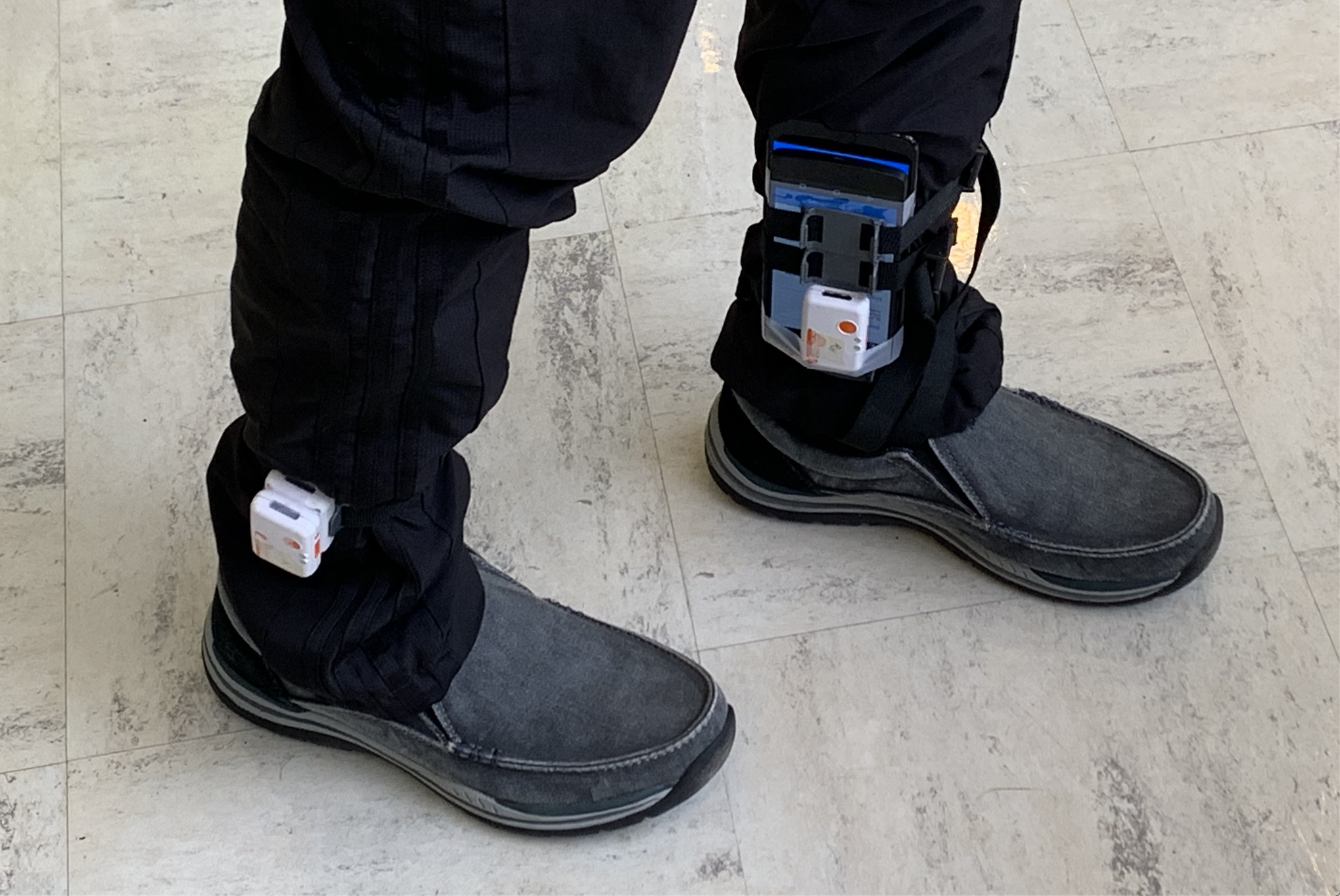}
    \caption{Sensor configuration. Devices are placed on top of each other vertically.} 
    \label{fig:sensors}
\end{marginfigure}

To showcase this phenomena, we gathered accelerometers of different qualities to compare their performance. Four types of activity were performed -- walking, running, biking, and stair climbing -- in order to test the accelerometers' sensitivity to force. A total of five devices were placed on a participant's legs vertically on top of each other to correct for orientation and ensure the same amount of force is applied to each device\footnote{Accelerometers record data on XYZ planes with respect to the orientation of the device. If devices are not oriented the same direction, the signals are not directly comparable.}.
Three different devices were placed on the left leg while two of the same device were placed on the right leg to be used as a baseline measurement (see \autoref{fig:sensors}). This allows us to perform four comparisons per activity session recorded.

\begin{marginfigure}[1pc]
    \centering
    \includegraphics[width=0.98\linewidth]{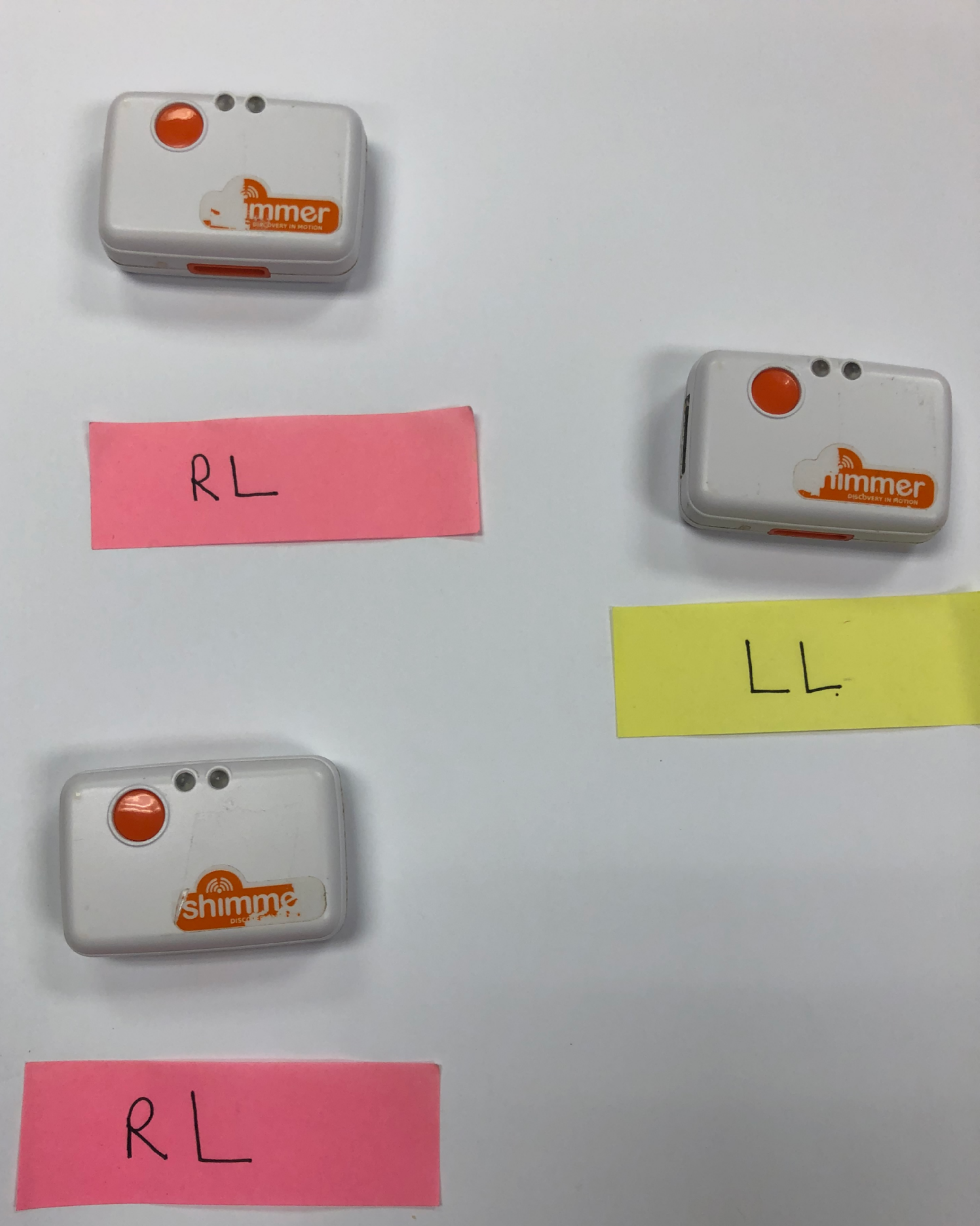}
    \caption{Shimmer3 sensors. These dedicated devices are one of the three devices use to record data. Two are strapped to right-leg (RL) and one on left-leg (LL). } 
    \label{fig:shimm}

\end{marginfigure}

\subsection{\textbf{Specification}}
For this study, the devices used to capture accelerometer data are as follows (see \autoref{tab:table2}):

    \underline{Inertial Measurement Unit (IMU):}
    An IMU is a portable electronic device to measure motion performed on a body. These can include acceleration, velocity, and orientation of the body, using a combination of:
    an accelerometer to measure acceleration for detecting sudden motion; a gyroscope to measure orientation and angular velocity; and a magnetometer to capture magnetic forces, acting as a compass.

    \begin{itemize}
        \item The \textit{Shimmer3} (see \autoref{fig:shimm}) is a portable IMU manufactured by Shimmer Sensing Inc.\footnote{http://www.shimmersensing.com/} that records accelerometer, gyroscope, and magnetometer data on local storage or streamed over Bluetooth. Multiple Shimmer3 devices can be used together in a single session, with automatic synchronization. For this study, data was stored locally and retrieved later.
    \end{itemize}

\underline{Smartphone:}
Many of the features we come to expect from our phones are only possible thanks to the same sensors used in IMUs. A phone's ability to switch between landscape and portrait, or wake when picked up, are good examples of how an accelerometer is used on modern phones. An application called ``Accelerometer Data Recorder''\footnote{https://play.google.com/store/apps/details?id=pt.acoelhosantos. android.acc} was used to collect data from the following devices.

\begin{itemize}

    \item The \textit{Motorola Z2 Force} is an Android smartphone released in 2017. Of the devices used in this case study, this is the most recent. 

    \item The \textit{Nexus 5} is an Android smartphone manufactured by LG for Google, released in 2013. Of the devices used in this case study, this is the oldest.

\end{itemize}

\subsection{\textbf{What We Did?}}

\begin{figure*}[!t]
  \centering

    \begin{minipage}[m]{0pt}
    \hspace{-150pt}
        \vspace{150pt}
    \fbox{
    \begin{minipage}{125pt}
      \textbf{Accelerometer Visualization}
      
      \vspace{1pc} The visualization in \autoref{fig:rawsignal} shows 170 seconds of activity across three devices. Each monitors the XYZ axis respective to each device and are shown with the following colors: blue \textcolor{blue}{x-axis}, red \textcolor{red}{y-axis}, green \textcolor{green}{z-axis}. 
      
      \vspace{1pc} In order to compare devices, a synchronized clock is required, which was done as pre-processing using this visualization. The effectiveness of this approach is limited to shorter sequences.
    \end{minipage}}
    \end{minipage}
    \begin{minipage}[m]{0.99\linewidth}
         \includegraphics[height=390pt]{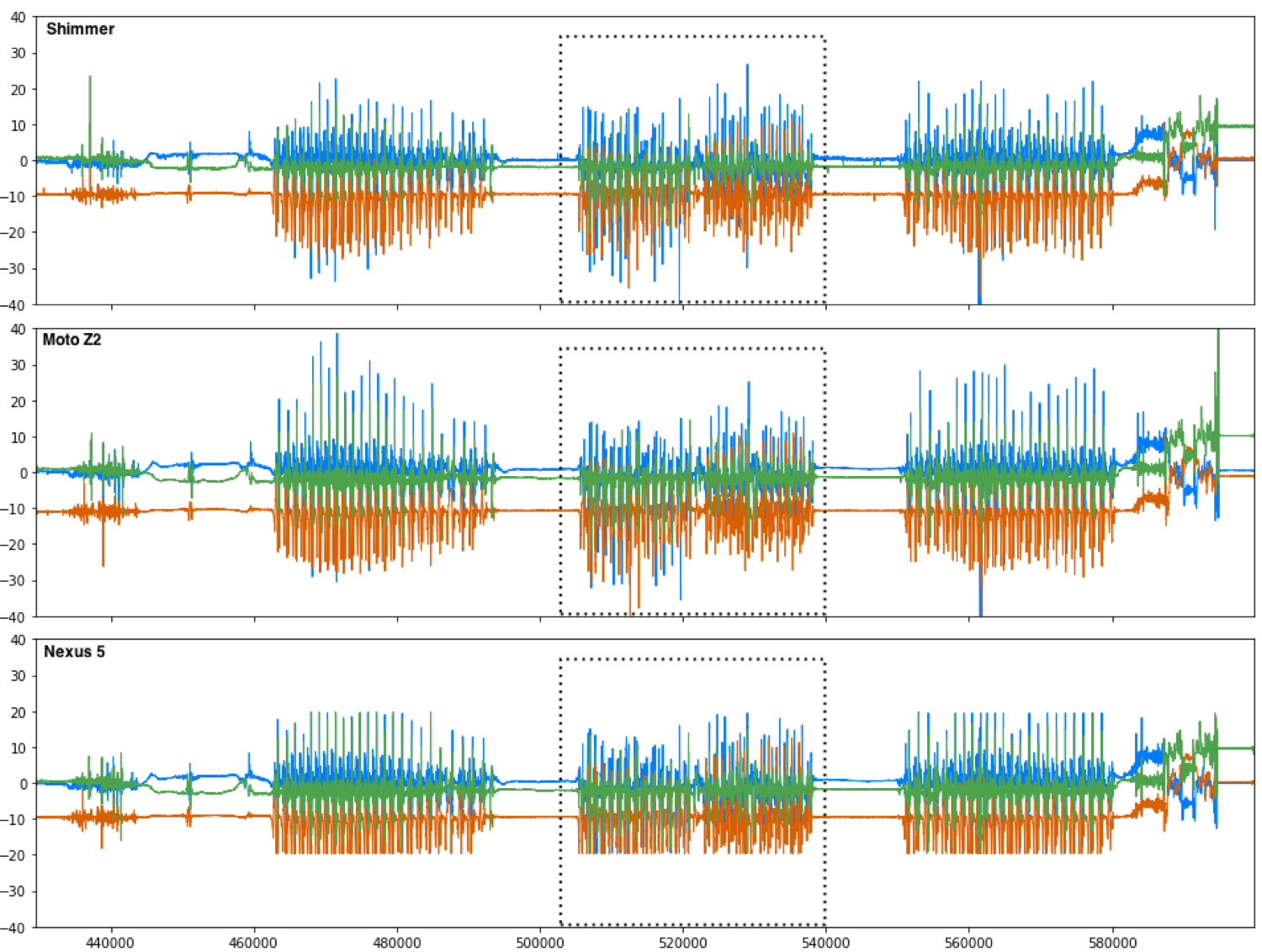}
    \end{minipage}
  
  \caption{This visualization represents accelerometer data for an activity (staircase, left leg) for a participant wearing three devices (top left). Signals show pre and post activity data from accelerometers. The highlighted area is the target activity for 60 seconds.}
  \label{fig:rawsignal}
\end{figure*}

For this case study, we recorded data from four participants for each of the four activities, for a total of 16 sessions. Each session is comprised of five devices---left leg: Shimmer3, Moto Z2, Nexus 5; right leg: Shimmer3, Shimmer3  (see \autoref{fig:sensors})---recording data from three axes (XYZ). The total amount of time the participants were wearing the accelerometers varies but the actual activity recording time was 60 seconds. Each of the data streams contained additional forces measured before and after the target activity (see \autoref{fig:rawsignal}).

\begin{figure*}[!t]
  \centering
  
    \begin{minipage}[m]{0pt}
    \hspace{-150pt}
        \vspace{150pt}
    \fbox{
    \begin{minipage}{125pt}
      \textbf{Target Activity Visualization}
      
      \vspace{1pc} The zoomed raw signal visualization in \autoref{fig:zoomedsignal}  shows 60 seconds of staircase climbing (climbing up and down is separated by a gap). The arrows marked represent significant variations in the detected forces, whose identification is the objective of our investigation. 

      \vspace{1pc}Here we can observe differences in the magnitudes recorded when the device is in motion and signal jitter when the device is stationary.
      \end{minipage}}
    \end{minipage}
    \begin{minipage}[m]{0.99\linewidth}
         \includegraphics[height=390pt]{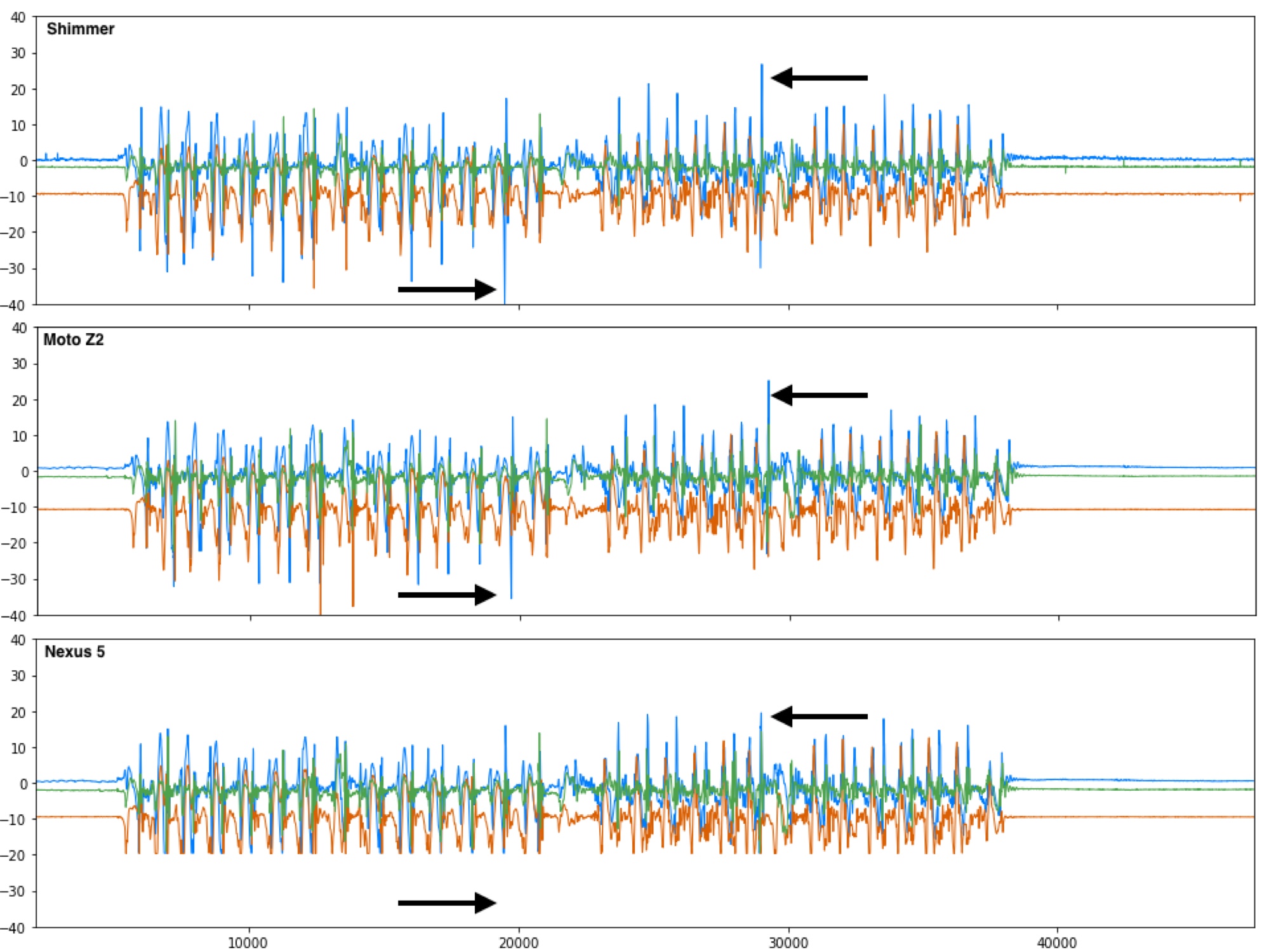}
    \end{minipage}
  
  \caption{This visualization represents accelerometer data for an activity (staircase, right leg) for a participant wearing three devices (top left). Signals show pre and post activity data from accelerometers. The highlighted area is the target activity for 60 seconds.}
  \label{fig:zoomedsignal}
\end{figure*}

Once all recordings were obtained from the participants, the signals were aligned and cropped to allow the same length of time to be evaluated. Not all devices can be started and stopped at the same time, so the portions of time where not all devices were recording were removed. Alignment was meant to be performed by tagging each data point with Unix timestamps, but each device had small variations in their clock which had to be corrected. This was performed visually, as auto-correlation methods were not able to align the signals properly. Each of the devices sampled at a different rate, so special attention was given to aligning based on time and not point position in a data array.

%% file: Findings.tex
\begin{marginfigure}[8pc]
  \begin{minipage}{0.98\linewidth}
    \centering
    \includegraphics[width=\linewidth]{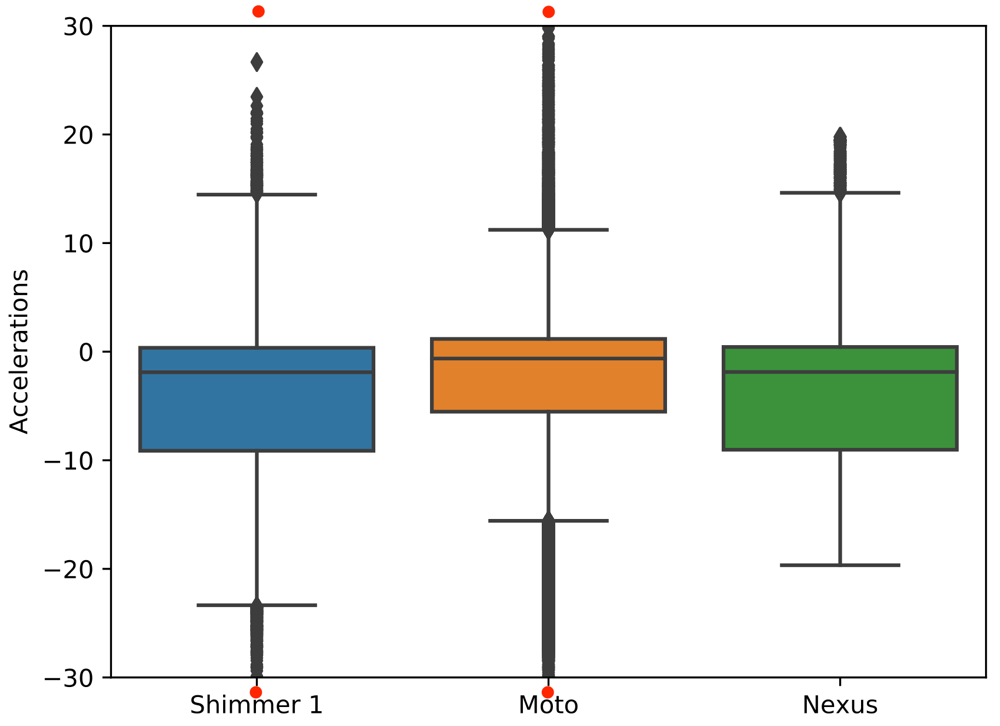}
    \caption{Boxplot of inter-device variation (staircase, left leg).}
    \label{fig:boxplot_ll}
  \end{minipage}
\end{marginfigure}

\begin{marginfigure}[0pc]
  \begin{minipage}{0.98\linewidth}
    \centering
    \includegraphics[width=\linewidth]{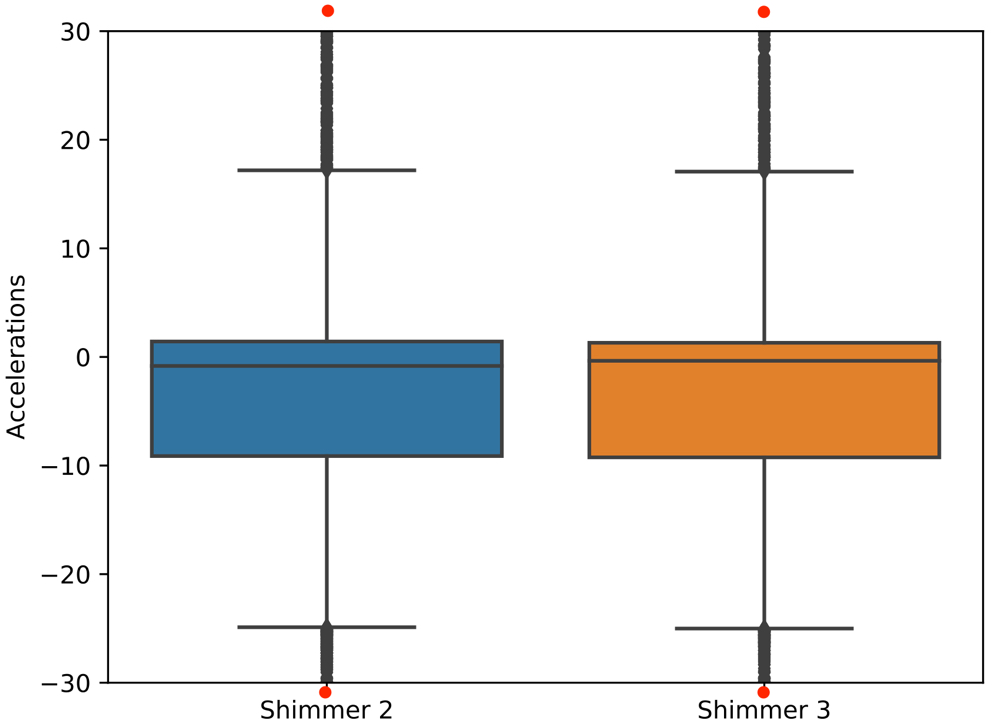}
    \caption{Boxplot of Shimmer-to-Shimmer variation (staircase, right leg).}
    \label{fig:boxplot_rl}
  \end{minipage}
\end{marginfigure}

The objective of this case study was to understand the variability of sensors between different devices.
Accelerometer data is complex and multivariate, and dealing with measurements from sensors needs intuitive visualization to help identify irregularities in the data~\cite{richter2009visualizing}.
From the visualizations showing the complete (see \autoref{fig:rawsignal}) and zoomed signals (see \autoref{fig:zoomedsignal}), we can observe that some peaks are more pronounced between devices at the same timestamp. Additionally, some of the devices consistently showed more jitter, which could be either small forces or noise.

The variations in the signal ranges were captured using boxplots (see Figures \ref{fig:boxplot_ll} and \ref{fig:boxplot_rl}) that show the variation in the means and standard deviations within the same session of recording. From these we can see the devices contain different means, standard deviations, and outliers, but it is not clear if the difference in calculations are a result of noise or device differences. 
One of the reasons behind the differences could relate to the limits of how much force a sensor can capture. From the four types of devices used, only one can observe 160$m/s^{2}$ force of acceleration. Data points captured at high accelerations can skew the mean and standard deviation values.

\begin{figure}[!bt]
  \centering
  \frame {\includegraphics[width=0.95\columnwidth]{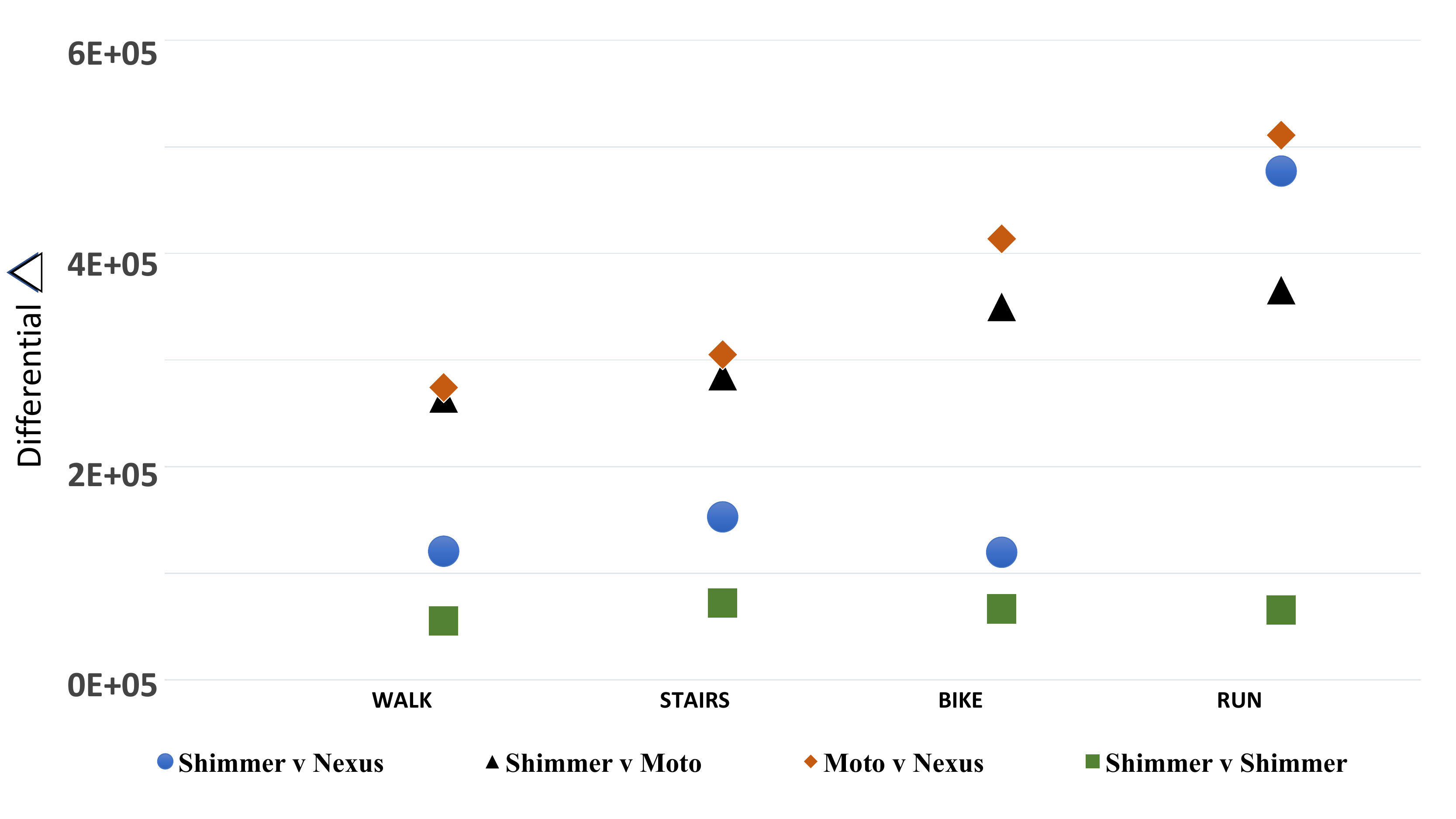}}
  \caption{Dynamic Time Warping (DTW) comparison between devices and activities. Each symbol represents an average DTW for a device pair in each activity. The values represent the differentials between the signals in the device pair.}~\label{fig:cats}
\end{figure}

This is not necessarily the only reason for these differences, so we used Dynamic Time Warping (DTW)~\cite{cuesta2016practical} as an alternate method to compare how similar 2 signals are to each other. 
To do this, each signal was first down-sampled to 300Hz by using linear interpolation of the nearest samples in time.
With a total of 240 signals ( 4 [participants] $\times$ 4 [activities] $\times$ 5 [devices] $\times$ 3 [axis] ), we produced 960 DTW outcomes by performing four comparisons between pair of devices---Shimmer vs. Shimmer (right leg); Shimmer vs Moto Z2 (left leg); Shimmer vs Nexus 5 (left leg); Moto Z2 vs Nexus 5 (left leg). These values were averaged for every comparison at every activity and shown in \autoref{fig:cats}.

The resulting graph represents the differentials between the devices for each activity. Additionally, activities are sorted by the amount of force used. Most differentials increase as more intensive activities are taking place, which implies that the devices are producing signals that are deviating from each other. The only exception is our baseline (Shimmer vs Shimmer), where the variations remain consistent, though non-zero, over all activities. We can attribute the non-zero differential to noise or the device not being at the exactly identical location. 
In the other comparisons, the differential grew as more intensive activities occurred. This could be in part due to device limitation. For example, the largest differential was the Nexus device, which was limited to 20$m/s^{2}$ forces to acceleration.

%% file: Discussion.tex
The performance of wearable sensors can vary with the underlying activity they are attempting to monitor. If a researcher wants to reach a conclusion while using multiple devices as data sources, they mus take into account that the data obtained can contain variability even if it is from the same sensor, activity and person. It is up to the researcher to determine how much uncertainty is acceptable for the conclusion they're trying to reach. 

Additionally, it is not sufficient to simply evaluate specifications from a sensor maker as the devices attached to the sensor can also play a role in their performance. For example, the sampling rate achieved from the accelerometers on our Android devices was determined by the phone operating system and appeared limited to the processor speed.

This shows that variability in the data captured by sensors exist even on devices that are expected to be the same. Although this case study is limited in the amount of devices used. We can show the usefulness in having users gather initial data about their devices, specifications, performance and variability from a pilot studies before committing sensors for their experiments.

%% file: Conclusion.tex
The introduction of sensors used to capture data of our environment increases every year. Sensors such as accelerometers can be found in multiple scenarios, such as fitness bands, smartphones, and medical devices, but there is an inherent possibility of differences in measurements gathered from them. We investigated through a case-study four activities on four participants using three devices with accelerometers to identify variance between them. Through signal visualizations and dynamic time warping techniques we demonstrated the variance in measurement for the same type of accelerometers on different devices. Lastly, we suggest, as part of the research processes, researchers and users should perform pilot studies to evaluate their equipment or allow for considerable amount of variance in their data.

Though we performed a case study on accelerometers, we will continue this work on gyroscope and magnetometers. We will also suggest investigating and modelling the uncertainty in the data measurement on these sensors.